\documentclass[article,nojss]{jss}

\usepackage{orcidlink,thumbpdf,lmodern}
\usepackage{amsmath}
\usepackage{xcolor}

\newcommand{\class}[1]{`\code{#1}'}
\newcommand{\fct}[1]{\code{#1()}}

\author{Alberto Fern{\'a}ndez~\orcidlink{0000-0002-1241-1646}\\Universitat
  Rovira i Virgili
  \And Sergio G{\'o}mez~\orcidlink{0000-0003-1820-0062}\\Universitat Rovira i
  Virgili}
\Plainauthor{Alberto Fern{\'a}ndez, Sergio G{\'o}mez}

\title{\pkg{mdendro}: An \proglang{R} Package for Extended Agglomerative
  Hierarchical Clustering}
\Plaintitle{mdendro: An R Package for Extended Agglomerative Hierarchical
  Clustering}
\Shorttitle{\pkg{mdendro}: Extended Agglomerative Hierarchical Clustering in
  \proglang{R}}

\Abstract{
  \pkg{mdendro} is an \proglang{R} package that provides a comprehensive
  collection of linkage methods for agglomerative hierarchical clustering on a
  matrix of proximity data (distances or similarities), returning a
  multifurcated dendrogram or multidendrogram. Multidendrograms can group more
  than two clusters at the same time, solving the nonuniqueness problem that
  arises when there are ties in the data. This problem causes that different
  binary dendrograms are possible depending both on the order of the input data
  and on the criterion used to break ties. Weighted and unweighted versions of
  the most common linkage methods are included in the package, which also
  implements two parametric linkage methods. In addition, package \pkg{mdendro}
  provides five descriptive measures to analyze the resulting dendrograms:
  cophenetic correlation coefficient, space distortion ratio, agglomerative
  coefficient, chaining coefficient and tree balance.
}

\Keywords{multifurcated dendrogram, parametric linkage, dendrogram descriptor,
\proglang{R}}
\Plainkeywords{multifurcated dendrogram, parametric linkage,
dendrogram descriptor, R}

\Address{
  Alberto Fern{\'a}ndez\\
  Departament d'Enginyeria Qu{\'{\i}}mica\\
  Universitat Rovira i Virgili\\
  Av.\ Pa{\"{\i}}sos Catalans 26\\
  43007 Tarragona, Spain\\
  E-mail: \email{alberto.fernandez@urv.cat}\\
  \\
  Sergio G{\'o}mez\\
  Departament d'Enginyeria Inform{\`a}tica i Matem{\`a}tiques\\
  Universitat Rovira i Virgili\\
  Av.\ Pa{\"{\i}}sos Catalans 26\\
  43007 Tarragona, Spain\\
  E-mail: \email{sergio.gomez@urv.cat}\\
  URL: \url{https://webs-deim.urv.cat/~sergio.gomez/}
}

\usepackage{Sweave}
\begin{document}
\Sconcordance{concordance:mdendro.tex:mdendro.Rnw:1 13 1 1 6 55 1 1 0 119 1 %
1 16 14 1 1 2 7 0 1 2 21 1 1 13 1 2 51 1 1 5 7 0 1 2 6 1 1 2 1 %
0 10 1 3 0 1 2 2 1 1 12 1 2 29 1 1 2 1 0 11 1 3 0 1 2 2 1 1 13 %
1 2 30 1 1 2 1 0 8 1 1 2 4 0 1 2 2 1 1 12 1 2 67 1 1 2 1 0 1 1 %
1 4 6 0 1 2 2 1 1 7 1 2 34 1 1 2 1 0 1 1 19 0 1 2 5 1 1 2 1 0 %
2 1 3 0 1 2 1 1 1 2 1 0 4 1 3 0 1 2 2 1 1 2 7 0 1 2 1 1 1 2 7 %
0 1 2 20 1 1 2 1 0 1 1 1 4 3 0 1 1 1 4 6 0 1 2 2 1 1 12 1 2 35 %
1 1 2 1 0 2 1 1 4 6 0 1 2 2 1 1 8 1 2 54 1 1 2 1 0 2 1 1 4 6 0 %
1 2 2 1 1 8 1 2 83 1 1 5 4 0 4 1 1 4 6 0 1 2 2 1 1 13 1 2 50 1 %
1 2 1 0 6 1 3 0 1 2 2 1 1 8 1 2 8 1 1 2 1 0 1 1 5 0 2 1 6 0 1 %
2 4 1 1 2 1 0 1 1 5 0 1 1 6 0 1 2 15 1 1 2 1 0 5 1 3 0 1 2 2 1 %
1 7 1 2 12 1 1 20 1 2 14 1 1 2 1 0 3 1 1 3 2 0 3 1 1 5 7 0 1 2 %
2 1 1 16 1 2 7 1 1 2 4 0 1 2 2 1 2 2 11 1 1 2 1 0 2 1 3 0 1 2 %
2 1 1 4 1 2 93 1}


\section{Introduction}

Agglomerative hierarchical clustering (AHC) is widely used to classify
individuals into a hierarchy of clusters organized in a tree structure called
dendrogram \citep{Gordon:1999}. There are different types of AHC linkage
methods, such as single linkage, complete linkage, average linkage and Ward's
method, which only differ in the definition of the distance measure between
clusters. All these methods start from a distance matrix between individuals,
each one forming a singleton cluster, and gather clusters into groups of
clusters, this process being repeated until a complete hierarchy of partitions
into clusters is formed.

Except for the single linkage case, all the other AHC linkage methods suffer
from a nonuniqueness problem known as the ties in proximity problem. This
problem arises whenever there are more than two clusters separated by the same
minimum distance during the agglomerative process of a pair-group AHC algorithm.
This type of algorithm breaks ties choosing any pair of clusters, and proceeds
in the same way until a binary dendrogram is obtained. However, different binary
dendrograms are possible depending both on the order of the input data and on
the criterion used to break ties.

The ties in proximity problem is long known \citep{Hart:1983, Morgan+Ray:1995,
Backeljau+al:1996}, even from studies in different fields, such as biology
\citep{Arnau+Mars+Marin:2005}, psychology \citep{VanDerKloot+Spaans+Heiser:2005}
and chemistry \citep{MacCuish+Nicolaou+MacCuish:2001}. The extend of the problem
in a particular field has been analyzed for microsatellite markers
\citep{Segura+al:2022}. Nevertheless, this problem is ignored by some software
packages: function \fct{hclust} in package \pkg{stats} and function \fct{agnes}
in package \pkg{cluster} of \proglang{R} \citep{R:2021}, commands \fct{cluster}
and \fct{clustermat} of \proglang{Stata} \citep{Stata:2021}, function
\fct{linkage} in the \pkg{Statistics and Machine Learning Toolbox} of
\proglang{MATLAB} \citep{MATLAB:2022}, and function \fct{hclust} in package
\pkg{Clustering.jl} of \proglang{Julia} \citep{Julia:2017}.

There are some other statistical packages that just warn against the existence
of the nonuniqueness problem in AHC. For instance, procedure \code{Hierarchical
Cluster Analysis} of \proglang{SPSS Statistics} \citep{SPSS:2021}, procedure
\code{CLUSTER} of \proglang{SAS} \citep{SAS:2018}, function \fct{Agglomerate} in
the \pkg{Hierarchical Clustering Package} of \proglang{Mathematica}
\citep{Mathematica:2020}, and also function \fct{linkage} in module
\code{scipy.cluster.hierarchy} of package \pkg{SciPy} in \proglang{Python}
\citep{SciPy:2020}.

Software packages that do not ignore the nonuniqueness problem fail to adopt a
common standard with respect to ties, and they simply break ties in any
arbitrary way. Here we introduce \pkg{mdendro}, an \proglang{R} package that
implements a variable-group AHC algorithm \citep{Fernandez+Gomez:2008} to solve
the nonuniqueness problem found in any pair-group AHC algorithm.

Package \pkg{mdendro} was designed using state-of-the-art methods based on
neighbor chains, and its base code was implemented in \proglang{C++}. It was
developed using object-oriented programming, where each linkage method
constitutes a different class. This eases the extensibility of the package since
other linkage methods can be added as new classes. The package is available from
the Comprehensive \proglang{R} Archive Network (CRAN) at
\url{https://CRAN.R-project.org/package=mdendro} and on GitHub at
\url{https://github.com/sergio-gomez/mdendro}. The functionality of the
\proglang{R} package \pkg{mdendro} makes it very similar and compatible with the
main ones currently in use, namely the \proglang{R} functions \fct{hclust} in
package \pkg{stats} and \fct{agnes} in package \pkg{cluster}
\citep{Maechler+al:2021}. The result is a package \pkg{mdendro} that includes
and extends the functionality of these reference functions.

The rest of the article is structured as follows. In
Section~\ref{sec:algorithms} we describe the pair-group and the variable-group
AHC algorithms, the latter grouping more than two clusters at the same time when
ties occur. Section~\ref{sec:methods} describes the most common AHC linkage
methods: single linkage, complete linkage, average linkage, centroid linkage and
Ward's method. Package \pkg{mdendro} also includes two parametric linkage
methods: $\beta$-flexible linkage and versatile linkage. In the same section,
five descriptive measures for the resulting dendrograms are included:
cophenetic correlation coefficient, space distortion ratio, agglomerative
coefficient, chaining coefficient and tree balance. Section~\ref{sec:comparison}
compares package \pkg{mdendro} with other state-of-the-art packages for AHC.
Finally, in Section~\ref{sec:summary}, we give some concluding remarks.


\section{Agglomerative hierarchical clustering algorithms}
\label{sec:algorithms}

\subsection{Pair-group algorithm}

AHC algorithms build a hierarchical tree in a bottom-up way, from a matrix of
pairwise distances between individuals of a set $\Omega=\{x_{1},\ldots,x_{n}\}$.
The pair-group algorithm \citep{Sneath+Sokal:1973} has the following steps:
\begin{enumerate}
  \item[0)] Initialize $n$ singleton clusters with one individual in each one of
    them: $X_{1}=\{x_{1}\}$, \ldots, $X_{n}=\{x_{n}\}$. Initialize also the
    distances between clusters, $D(X_{i},X_{j})$, with the values of the
    distances between individuals, $d(x_{i},x_{j})$:
    \begin{displaymath}
      D(X_{i},X_{j}) = d(x_{i},x_{j}) \, , \qquad \forall i,j=1,\ldots,n
        \, .
    \end{displaymath}
  \item[1)] Find the shortest distance separating two different clusters,
    $D_{\mathrm{shortest}}$.
  \item[2)] Select two clusters $X_{i}$ and $X_{i'}$ separated by the shortest
    distance $D_{\mathrm{shortest}}$, and merge them into a new cluster
    $X_{i} \cup X_{i'}$.
  \item[3)] Compute the distances $D(X_{i} \cup X_{i'},X_{j})$ between the new
    cluster $X_{i} \cup X_{i'}$ and each one of the other clusters $X_{j}$.
  \item[4)] If all the individuals are not in the same cluster yet, then go back
    to step~1.
\end{enumerate}

The nonuniqueness problem in the pair-group algorithm arises when two or more
shortest distances between different clusters are equal during the agglomerative
process \citep{Hart:1983}. The standard approach consists in choosing only a
single pair to break the tie. However, different hierarchical clusterings are
possible depending on the criterion used to break ties (usually a pair is just
chosen at random), and the user is unaware of this problem.

\begin{figure}[t!]
\centering
\includegraphics{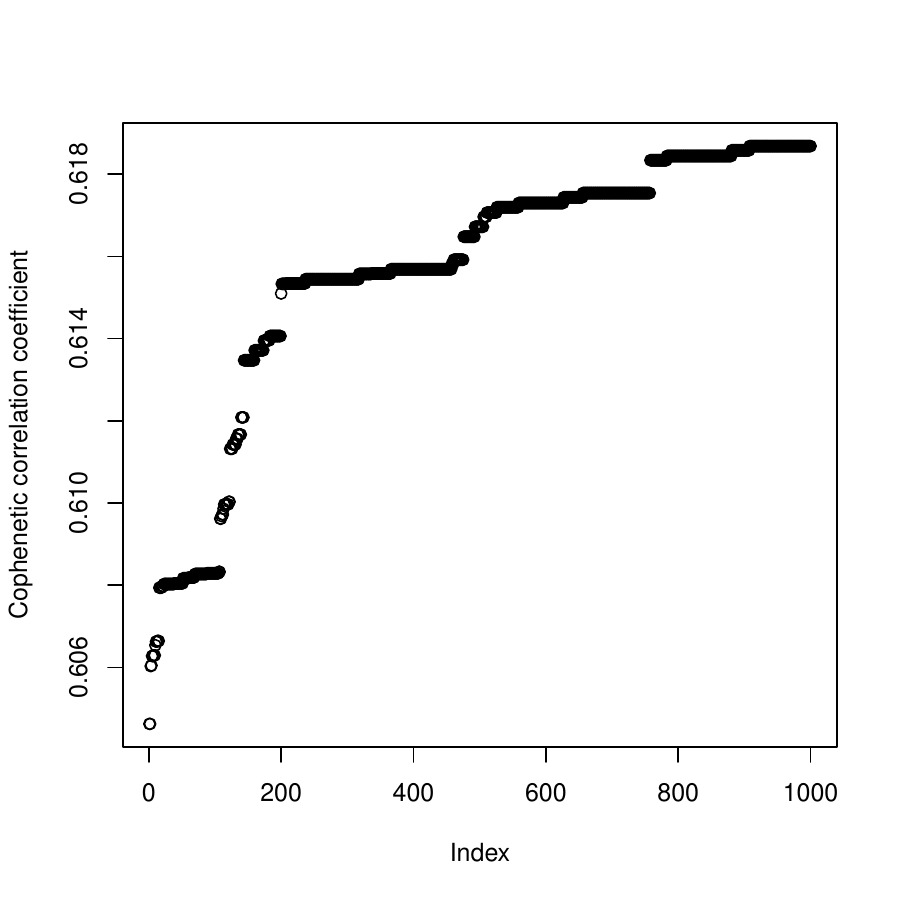}
\caption{\label{fig:permutations} Sorted cophenetic correlation coefficients for
the different pair-group dendrograms obtained by using random permutations of
the same grapevine cultivars dataset.}
\end{figure}

For example, let us consider the genetic profiles of 51 grapevine cultivars at
six microsatellite loci \citep{Almadanim+al:2007}. Microsatellites are a type of
molecular markers and, as such, they are useful to characterize genotypes and to
study genetic diversity within and between species. The distance between
genotypes of two grapevine cultivars is defined, using microsatellite markers,
as one minus the fraction of shared alleles, and this definition is used here to
calculate a distance matrix \code{d}. Hierarchical clustering of microsatellites
is prone to generate tied distances because the number of shared alleles can
only take values between zero and the total number of alleles, which is usually
a small number \citep{Segura+al:2022}. In this example with grapevine cultivars,
where there are just six microsatellite loci, the number of pairs of genotypes
separated by the same distance is relatively large and, consequently, there are
very few different distances:
\begin{Schunk}
\begin{Sinput}
R> length(unique(d))
\end{Sinput}
\begin{Soutput}
[1] 11
\end{Soutput}
\end{Schunk}
As a consequence of these 11 unique values out of 1275 pairwise distances in the
matrix, it becomes very easy to find tied distances during the agglomeration
process. The reach of the nonuniqueness problem for this example is the
existence of 11,160 structurally different binary dendrograms. This number
corresponds to the average linkage method and a resolution of 3 decimal digits,
and it has been computed using the \code{Hierarchical_Clustering} tool in
\pkg{Radatools} \citep{Gomez+Fernandez:2021}. We can check the diversity of
results by just calculating binary dendrograms for random permutations of the
data and plotting the range of values of their cophenetic correlation
coefficients (see Figure~\ref{fig:permutations}), what clearly indicates the
existence of many structurally different binary dendrograms. In cases like this,
where different dendrograms are possible, the reproducibility of results is
compromised because, depending on the input order of data \citep{Podani:1997},
or depending on the particular implementation of the pair-group algorithm used,
different results may be obtained. The interpretation of these results may be
biased towards just one of the different solutions, and any conclusion drawn
from a single dendrogram must be considered partial and, therefore,
questionable.

\subsection{Variable-group algorithm}

\citet{Fernandez+Gomez:2008} introduced a variable-group algorithm to ensure
uniqueness in AHC, which differs from the pair-group algorithm in the following
steps:
\begin{enumerate}
  \item[2)] Select all the groups of clusters separated by the shortest distance
    $D_{\mathrm{shortest}}$, and merge them into several new clusters
    $X_{I}=\bigcup_{i \in I}X_{i}$, each one made up of several
    subclusters $X_{i}$ indexed by $i$ in $I=\{i_{1},\ldots,i_{p}\}$.
  \item[3)] Compute the distances $D(X_{I},X_{J})$ between any two clusters
    $X_{I}=\bigcup_{i \in I}X_{i}$ and $X_{J}=\bigcup_{j \in J}X_{j}$, each one
    of them made up of several subclusters $X_{i}$ and $X_{j}$ indexed by $i$ in
    $I=\{i_{1},\ldots,i_{p}\}$ and $j$ in $J=\{j_{1},\ldots,j_{q}\}$,
    respectively.
\end{enumerate}

When there are tied shortest distances in the agglomerative process, in order to
keep track of valuable information regarding the heterogeneity of the clusters
that are formed, function \fct{linkage} in package \pkg{mdendro} saves a fusion
interval $[D_{\min}(X_{I}),D_{\max}(X_{I})]$ for each cluster $X_{I}$ made up of
more than one subcluster ($|I|>1$), where:
\begin{eqnarray*}
  D_{\min}(X_{I}) = \min_{i \in I} \, \min_{\substack{i' \in I \\ i' \not = i}}
    \, D(X_{i},X_{i'}) \, , \\
  D_{\max}(X_{I}) = \max_{i \in I} \, \max_{\substack{i' \in I \\ i' \not = i}}
    \, D(X_{i},X_{i'}) \, .
\end{eqnarray*}

The variable-group algorithm groups more than two clusters at the same time when
ties occur, giving rise to a graphical representation called multidendrogram.
Its main properties are:
\begin{itemize}
  \item When there are no ties, the variable-group algorithm gives the same
    results as the pair-group one.
  \item It always gives a uniquely-determined solution.
  \item In the multidendrogram representation for the results, one can
    explicitly observe the occurrence of ties during the agglomerative process.
    Furthermore, the range of any fusion interval indicates the degree of
    heterogeneity inside the corresponding cluster.
\end{itemize}

\begin{figure}[t!]
\centering
\includegraphics{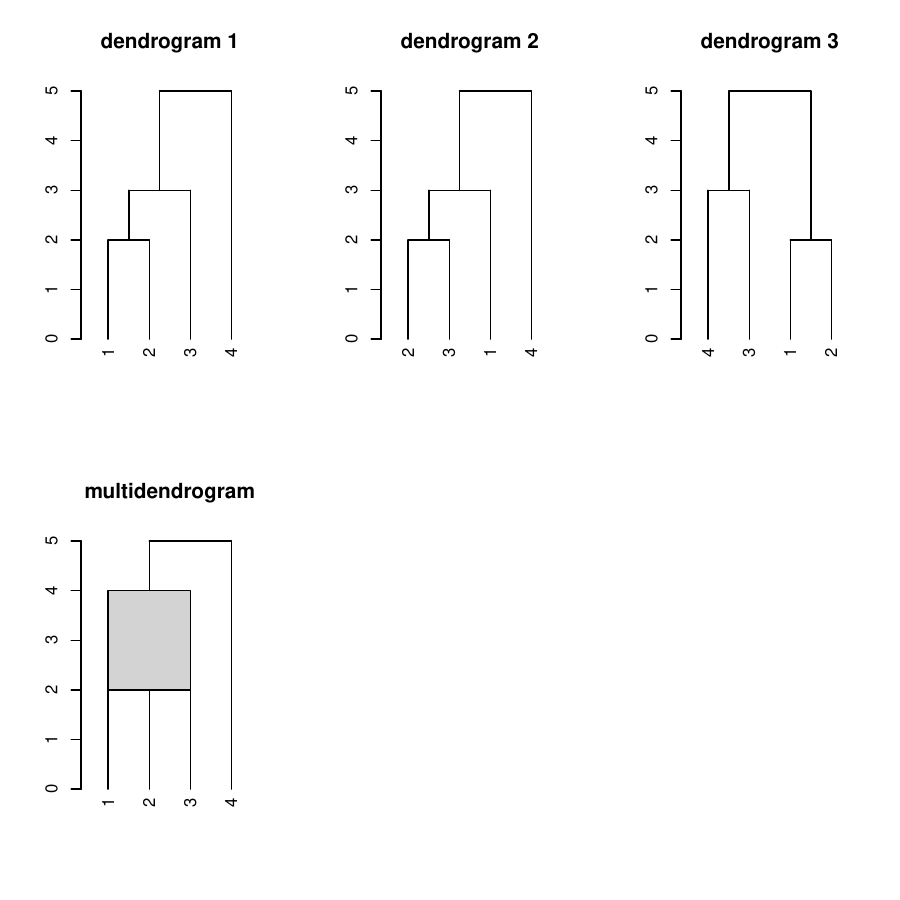}
\caption{\label{fig:toy} Different pair-group dendrograms and a unique
variable-group multidendrogram, all of them obtained using the average linkage
method. Observing the third dendrogram, one could get the wrong conclusion that
$x_{3}$ is closer to $x_{4}$ rather than to $x_{1}$ or $x_{2}$.}
\end{figure}

For example, let us suppose that we have a set of four individuals
$\{x_{1},x_{2},x_{3},x_{4}\}$, where the initial pairwise distances between them
are:
\begin{Schunk}
\begin{Sinput}
R> d <- as.dist(matrix(c(0, 2, 4, 7,
+                        2, 0, 2, 5,
+                        4, 2, 0, 3,
+                        7, 5, 3, 0), nrow = 4))
\end{Sinput}
\end{Schunk}
Notice that there are two pairs of individuals, $(x_{1},x_{2})$ and
$(x_{2},x_{3})$, separated by the shortest distance in the matrix, which is~2.
On the one hand, using the pair-group algorithm, we can obtain three different
binary dendrograms depending on the order of rows and columns in the distance
matrix (see Figure~\ref{fig:toy}):
\begin{Schunk}
\begin{Sinput}
R> par(mfrow = c(2, 3))
R> lnk1 <- linkage(d, group = "pair")
R> plot(lnk1, main = "dendrogram 1")
R> d2 <- as.dist(as.matrix(d)[c(2, 3, 4, 1), c(2, 3, 4, 1)])
R> lnk2 <- linkage(d2, group = "pair")
R> plot(lnk2, main = "dendrogram 2")
R> d3 <- as.dist(as.matrix(d)[c(4, 1, 2, 3), c(4, 1, 2, 3)])
R> lnk3 <- linkage(d3, group = "pair")
R> plot(lnk3, main = "dendrogram 3")
R> lnk4 <- linkage(d, group = "variable")
R> plot(lnk4, main = "multidendrogram")
\end{Sinput}
\end{Schunk}
On the other hand, using the variable-group algorithm, we obtain a unique
multidendrogram where $x_{1}$, $x_{2}$ and $x_{3}$ are grouped in a single
cluster at the same time. This new cluster is assigned two height values,
corresponding to the minimum and the maximum distances separating any two of the
constituent clusters $\{x_{1}\}$, $\{x_{2}\}$ and $\{x_{3}\}$. In this case, the
minimum distance is~2 and the maximum distance is~4 (see shadowed rectangle in
the multidendrogram of Figure~\ref{fig:toy}). Finally, the distance between the
new cluster $\{x_{1},x_{2},x_{3}\}$ and the cluster $\{x_{4}\}$ is calculated.
In case of the average linkage method, this distance is equal to~5, that is, the
arithmetic mean among the values~7, 5 and~3, corresponding respectively to the
distances $d(x_{1},x_{4})$, $d(x_{2},x_{4})$ and $d(x_{3},x_{4})$.

\begin{figure}[t!]
\centering
\includegraphics{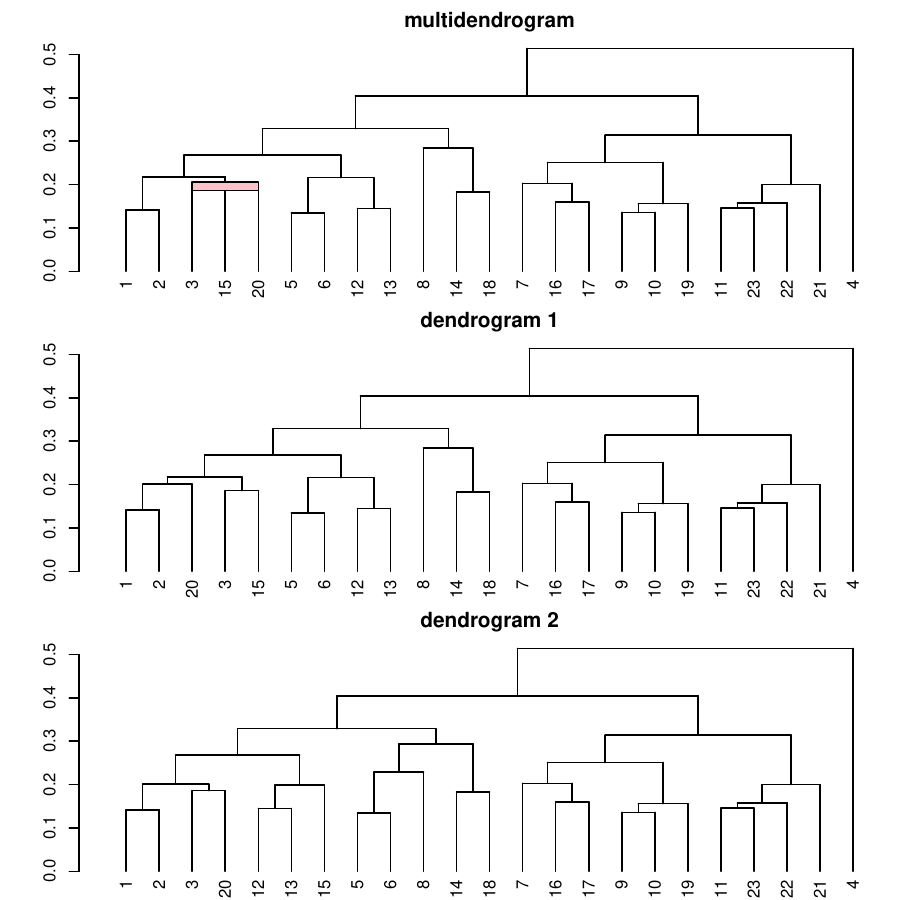}
\caption{\label{fig:soils} Complete linkage of the soils dataset yields a unique
variable-group multidendrogram and two different pair-group dendrograms. The two
pair-group dendrograms show important structural differences, despite simply arising
from how the tied distances are broken: selecting to join first soils~3 and~15 in
dendrogram~1, and soils~3 and~20 in dendrogram~2.}
\end{figure}

We show now one of the first examples in the literature that presented the ties
in proximity problem. This example corresponds to a study of 23 different soils,
whose similarity data are given in Table~2 of \cite{Morgan+Ray:1995}. As in the
original work, similarities are transformed into dissimilarities subtracting
them from 1, and the complete linkage method is used. The initial data present a
tied value for pairs of soils (3,15) and (3,20), which are responsible for the
existence of two different pair-group dendrograms; they are revealed by just
exchanging the positions of soils~15 and~20 in the distance matrix (see
Figure~\ref{fig:soils}):
\begin{Schunk}
\begin{Sinput}
R> soils <- 1 - as.dist(read.csv("soils.csv", header = FALSE))
R> s <- 1:23
R> attr(soils, "Labels") <- s
R> lnk <- linkage(soils, method = "complete", group = "variable")
R> lnk1 <- linkage(soils, method = "complete", group = "pair")
R> s[c(15, 20)] <- c(20, 15)
R> soils2 <- as.dist(as.matrix(soils)[s, s])
R> lnk2 <- linkage(soils2, method = "complete", group = "pair")
R> par(mfrow = c(3, 1), mar = c(1, 4, 2, 0))
R> plot(lnk, col.rng = "pink", main = "multidendrogram")
R> plot(lnk1, main = "dendrogram 1")
R> plot(lnk2, main = "dendrogram 2")
\end{Sinput}
\end{Schunk}
\cite{Morgan+Ray:1995} explain that the 23 soils have been categorized into
eight soil groups by a surveyor. Focusing on soils 1, 2, 6, 12 and 13, which are
the only members of the brown earths soil group, we can see that the second
pair-group dendrogram in Figure~\ref{fig:soils} does not place them in the same
cluster until they join soils from five other soil groups, forming the cluster
(1, 2, 3, 20, 12, 13, 15, 5, 6, 8, 14, 18). From this point of view, the
variable-group multidendrogram and the first pair-group dendrogram in
Figure~\ref{fig:soils} are better, since they keep soils 8, 14 and 18 out of the
brown earths soil group.

As we have already seen in the previous examples, with function \fct{linkage} we
can use both the pair-group algorithm and the variable-group one by setting the
\code{group} argument of function \fct{linkage} to either \code{"pair"} or
\code{"variable"}, respectively.

\begin{figure}[t!]
\centering
\includegraphics{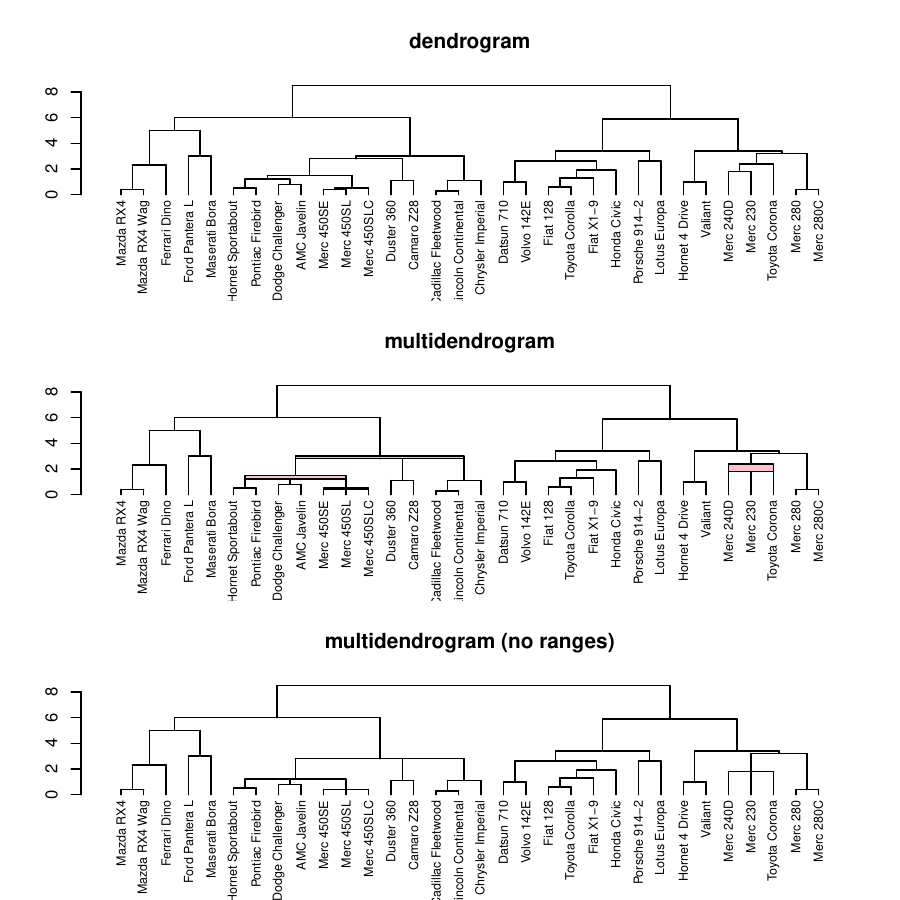}
\caption{\label{fig:ranges} Pair-group dendrogram vs.\ variable-group
multidendrogram. The ranges (rectangles) in the multidendrogram show the
heterogeneity of distances within the group, but they are optional in the plots
and can be hidden just by setting the \code{col.rng} argument in the
\fct{plot} function to \code{NULL}. Distances approximated to one decimal digit
both manually (top and center panels) and with the \code{digits} argument of
function \fct{linkage} (bottom panel).}
\end{figure}

The identification of ties requires the selection of the number of significant
digits in the working dataset. For example, if the original distances are
experimentally obtained with a resolution of three decimal digits, two distances
that differ in the sixth decimal digit should be considered as equal. In
function \fct{linkage}, the user can control this level of resolution by
adjusting its \code{digits} argument (see Figure~\ref{fig:ranges}):
\begin{Schunk}
\begin{Sinput}
R> par(mfrow = c(3, 1))
R> cars <- dist(scale(mtcars))
R> cars1 <- round(cars, digits = 1)
R> nodePar <- list(cex = 0, lab.cex = 0.7)
R> lnk1 <- linkage(cars1, method = "complete", group = "pair")
R> plot(lnk1, main = "dendrogram", nodePar = nodePar)
R> lnk2 <- linkage(cars1, method = "complete", group = "variable",)
R> plot(lnk2, col.rng = "pink", main = "multidendrogram", nodePar = nodePar)
R> lnk3 <- linkage(cars, method = "complete", group = "variable", digits = 1)
R> plot(lnk3, col.rng = NULL, main = "multidendrogram (no ranges)",
+       nodePar = nodePar)
\end{Sinput}
\end{Schunk}


\section{Linkage methods} \label{sec:methods}

\subsection{Common linkage methods}

During each iteration of the AHC algorithm, the distances $D(X_{I},X_{J})$ have
to be computed between any two clusters $X_{I}=\bigcup_{i \in I}X_{i}$ and
$X_{J}=\bigcup_{j \in J}X_{j}$, each one of them made up of several subclusters
$X_{i}$ and $X_{j}$ indexed by $i$ in $I=\{i_{1},\ldots,i_{p}\}$ and $j$ in
$J=\{j_{1},\ldots,j_{q}\}$, respectively. \citet{Lance+Williams:1966} introduced
a formula for integrating several AHC linkage methods into a single system,
avoiding the need of a separate computer program for each one of them.
Similarly, \citet{Fernandez+Gomez:2008} gave a variable-group generalization of
this formula, compatible with the fusion of more than two clusters
simultaneously:
\begin{equation}
  \label{eq:Lance_Williams}
  D(X_{I},X_{J}) = \sum_{i \in I} \sum_{j \in J} \alpha_{ij} D(X_{i},X_{j}) +
  \sum_{i \in I} \sum_{\substack{i' \in I\\i'>i}} \beta_{ii'} D(X_{i},X_{i'}) +
  \sum_{j \in J} \sum_{\substack{j' \in J\\j'>j}} \beta_{jj'} D(X_{j},X_{j'})\,.
\end{equation}
Function \fct{linkage} in package \pkg{mdendro} uses this recurrence relation to
compute the distance $D(X_{I},X_{J})$ from the distances $D(X_{i},X_{j})$
obtained during the previous iteration, being unnecessary to look back at the
initial distance matrix $d(x_{i},x_{j})$ at all. The values of the parameters
$\alpha_{ij}$, $\beta_{ii'}$ and $\beta_{jj'}$ determine the nature of the AHC
linkage methods \citep{Fernandez+Gomez:2008}. Some of these methods even have
weighted and unweighted forms, which differ in the weights assigned to
individuals and clusters during the agglomerative process: weighted methods
assign equal weights to clusters, while unweighted methods assign equal weights
to individuals. Package \pkg{mdendro} implements weighted and unweighted forms
of the most commonly used AHC linkage methods, namely:
\begin{itemize}
  \item \code{single}: the proximity between clusters equals the minimum
    distance or the maximum similarity between objects.
  \item \code{complete}: the proximity between clusters equals the maximum
    distance or the minimum similarity between objects.
  \item \code{arithmetic}: the proximity between clusters equals the arithmetic
    mean proximity between objects. Also known as average linkage, WPGMA
    (weighted version) or UPGMA (unweighted version).
  \item \code{ward}: the distance between clusters is a weighted squared
    Euclidean distance between the centroids of each cluster. This method is
    available only for distance data.
  \item \code{centroid}: the distance between clusters equals the square of the
    Euclidean distance between the centroids of each cluster. Also known as
    WPGMC (weighted version) or UPGMC (unweighted version). This method is
    available only for distance data. Note that both centroid versions, weighted
    and unweighted, may yield inversions that make dendrograms difficult to
    interpret.
\end{itemize}

\begin{figure}[t!]
\centering
\includegraphics{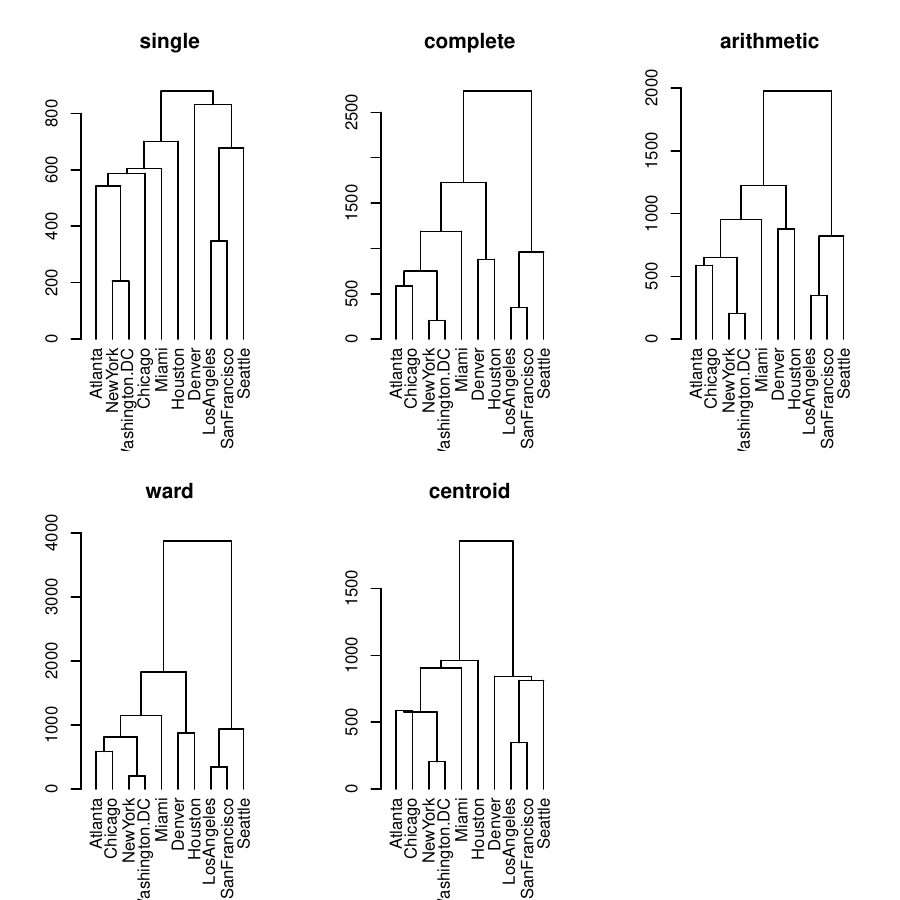}
\caption{\label{fig:methods} Common linkage methods on the \code{UScitiesD}
dataset.}
\end{figure}

In Figure~\ref{fig:methods}, we can see the differences between these AHC
linkage methods on the \code{UScitiesD} dataset, a matrix of distances between
a few US cities:
\begin{Schunk}
\begin{Sinput}
R> par(mfrow = c(2, 3))
R> methods <- c("single", "complete", "arithmetic", "ward", "centroid")
R> for (m in methods) {
+    lnk <- linkage(UScitiesD, method = m)
+    plot(lnk, cex = 0.6, main = m)
+  }
\end{Sinput}
\end{Schunk}

\subsection{Descriptive measures} \label{subsec:descriptors}

The result of function \fct{linkage} is an object of class \class{linkage} that
describes the resulting dendrogram. In particular, this object contains the
following calculated descriptors:
\begin{itemize}
  \item \code{cor}: Cophenetic correlation coefficient \citep{Sokal+Rohlf:1962},
  defined as the Pearson correlation coefficient between the output cophenetic
  proximity data and the input proximity data. It is a measure of how faithfully
  the dendrogram preserves the pairwise proximity between objects.
  \item \code{sdr}: Space distortion ratio \citep{Fernandez+Gomez:2020},
  calculated as the difference between the maximum and minimum cophenetic
  proximity data, divided by the difference between the maximum and minimum
  initial proximity data. Space dilation occurs when the space distortion ratio
  is greater than 1.
  \item \code{ac}: Agglomerative coefficient \citep{Rousseeuw:1986}, a number
  between 0 and 1 measuring the strength of the clustering structure obtained.
  \item \code{cc}: Chaining coefficient \citep{Williams+Lambert+Lance:1966}, a
  number between 0 and 1 measuring the tendency for clusters to grow by the
  addition of clusters much smaller rather than by fusion with other clusters of
  comparable size.
  \item \code{tb}: Tree balance \citep{Fernandez+Gomez:2020}, a number between 0
  and 1 measuring the equality in the number of leaves in the branches concerned
  at each fusion in the hierarchical tree.
\end{itemize}

For instance, when we use function \fct{linkage} to calculate the complete
linkage of the \code{UScitiesD} dataset, we obtain the following summary for the
resulting dendrogram:
\begin{Schunk}
\begin{Sinput}
R> lnk <- linkage(UScitiesD, method = "complete")
R> summary(lnk)
\end{Sinput}
\begin{Soutput}
Call:
linkage(prox = UScitiesD,
        type.prox = "distance",
        digits = 0,
        method = "complete",
        group = "variable")

Number of objects: 10

Binary dendrogram: TRUE

Descriptive measures:
      cor       sdr        ac        cc        tb
0.8077859 1.0000000 0.7738478 0.3055556 0.9316262
\end{Soutput}
\end{Schunk}

While multidendrograms are unique, users may obtain structurally different
pair-group dendrograms by just reordering the data. As a consequence,
descriptors are invariant to permutations for multidendrograms, but not for
pair-group dendrograms. Let us calculate a variable-group multidendrogram and a
pair-group dendrogram for the same data:
\begin{Schunk}
\begin{Sinput}
R> cars <- round(dist(scale(mtcars)), digits = 1)
R> lnk1 <- linkage(cars, method = "complete", group = "variable")
R> lnk2 <- linkage(cars, method = "complete", group = "pair")
\end{Sinput}
\end{Schunk}
Now, if we apply a random permutation to data:
\begin{Schunk}
\begin{Sinput}
R> set.seed(1234)
R> ord <- sample(attr(cars, "Size"))
R> carsp <- as.dist(as.matrix(cars)[ord, ord])
R> lnk1p <- linkage(carsp, method = "complete", group = "variable")
R> lnk2p <- linkage(carsp, method = "complete", group = "pair")
\end{Sinput}
\end{Schunk}
We can check that the original and the permuted cophenetic correlation
coefficients are identical for variable-group multidendrograms:
\begin{Schunk}
\begin{Sinput}
R> c(lnk1$cor, lnk1p$cor)
\end{Sinput}
\begin{Soutput}
[1] 0.7782257 0.7782257
\end{Soutput}
\end{Schunk}
And they are different for pair-group dendrograms:
\begin{Schunk}
\begin{Sinput}
R> c(lnk2$cor, lnk2p$cor)
\end{Sinput}
\begin{Soutput}
[1] 0.7780010 0.7776569
\end{Soutput}
\end{Schunk}

\subsection{Parametric linkage methods}

\begin{figure}[t!]
\centering
\includegraphics{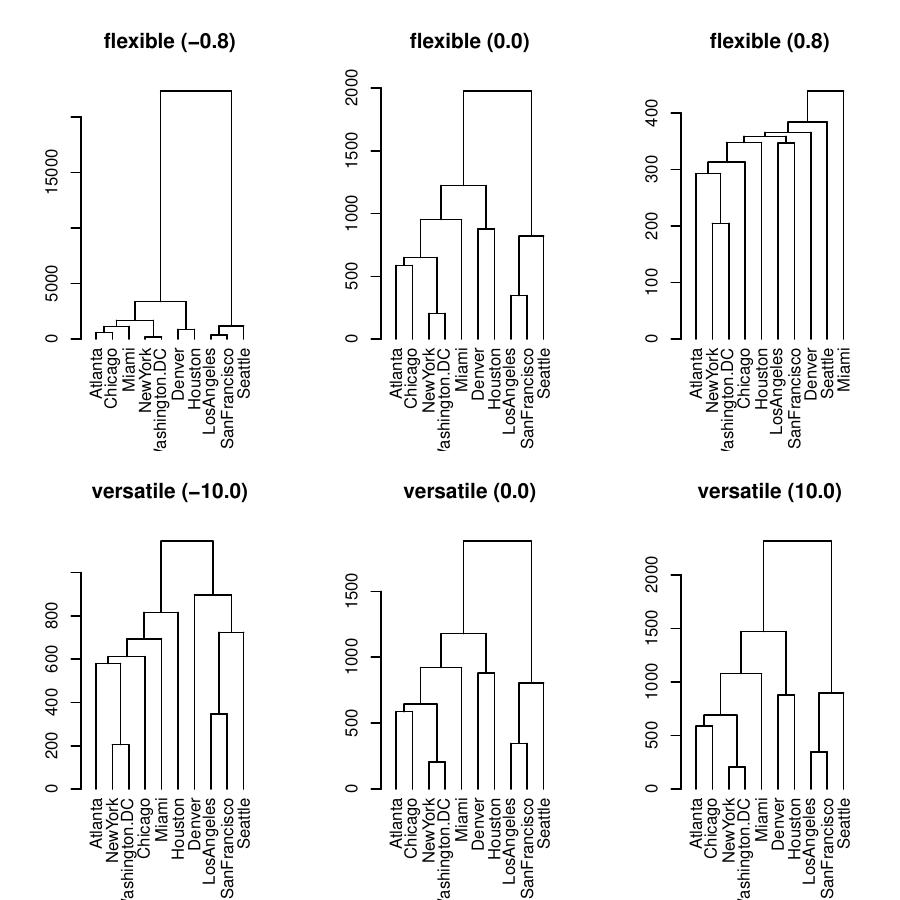}
\caption{\label{fig:parameter} Parametric linkage methods on the
\code{UScitiesD} dataset. Examples \code{flexible (0.8)} and \code{versatile
(-10.0)} are more space-contracting, while examples \code{flexible (-0.8)} and
\code{versatile (10.0)} are more space-dilating. Notice the huge heights of the
dendrogram \code{flexible (-0.8)}, taking into account that the maximum value of
the original distances is 2734.}
\end{figure}

Two of the AHC linkage methods available in package \pkg{mdendro},
\code{flexible} and \code{versatile}, depend on a parameter that takes values in
$[-1, +1]$ for \code{flexible} linkage, and in $(-\mathrm{Inf}, +\mathrm{Inf})$
for \code{versatile} linkage. In function \fct{linkage}, the desired value for
the parameter is passed through the \code{par.method} argument. This parameter
works as a cluster intensity coefficient, going from space-contracting
clustering strategies to space-dilating ones \citep{Fernandez+Gomez:2020}. On
the one hand, using \code{flexible} linkage with high values of the parameter or
\code{versatile} linkage with low values of the parameter, we obtain
space-contracting clusterings where dendrogram heights are shorter because
clusters move closer to other clusters as they grow. On the other hand, using
\code{flexible} linkage with low values of the parameter or \code{versatile}
linkage with high values of the parameter, we obtain space-dilating clusterings
where dendrogram heights are larger because clusters move further away from
other clusters as they grow. Here come some examples on the \code{UScitiesD}
dataset (see Figure~\ref{fig:parameter}):
\begin{Schunk}
\begin{Sinput}
R> par(mfrow = c(2, 3))
R> vals <- c(-0.8, 0.0, 0.8)
R> for (v in vals) {
+    lnk <- linkage(UScitiesD, method = "flexible", par.method = v)
+    plot(lnk, cex = 0.6, main = sprintf("flexible (%.1f)", v))
+  }
R> vals <- c(-10.0, 0.0, 10.0)
R> for (v in vals) {
+    lnk <- linkage(UScitiesD, method = "versatile", par.method = v)
+    plot(lnk, cex = 0.6, main = sprintf("versatile (%.1f)", v))
+  }
\end{Sinput}
\end{Schunk}

\subsubsection[Beta-flexible linkage]{$\beta$-flexible linkage}

Based on Equation~\ref{eq:Lance_Williams}, \citet{Lance+Williams:1967}
proposed an infinite system of AHC strategies defined by the following
constraint:
\begin{equation}
  \underbrace{\sum_{i \in I} \sum_{j \in J} \alpha_{ij}}_{\alpha} +
  \underbrace{\sum_{i \in I} \sum_{\substack{i' \in I \\ i'>i}} \beta_{ii'} +
              \sum_{j \in J} \sum_{\substack{j' \in J \\ j'>j}}
              \beta_{jj'}}_{\beta} = 1\,,
\end{equation}
where $-1 \le \beta \le +1$. Given a value of $\beta$, the value for
$\alpha_{ij}$ can be assigned following a weighted approach as in the original
$\beta$-flexible clustering method based on WPGMA and introduced by
\citet{Lance+Williams:1966}, or it can be assigned following an unweighted
approach as in the $\beta$-flexible clustering method based on UPGMA and
introduced by \citet{Belbin+Faith+Milligan:1992}. Further details can be
consulted in \citet{Fernandez+Gomez:2020}. When $\beta$ is set equal to~$0$,
\code{flexible} linkage is equivalent to \code{arithmetic} linkage.

\begin{figure}[t!]
\centering
\includegraphics{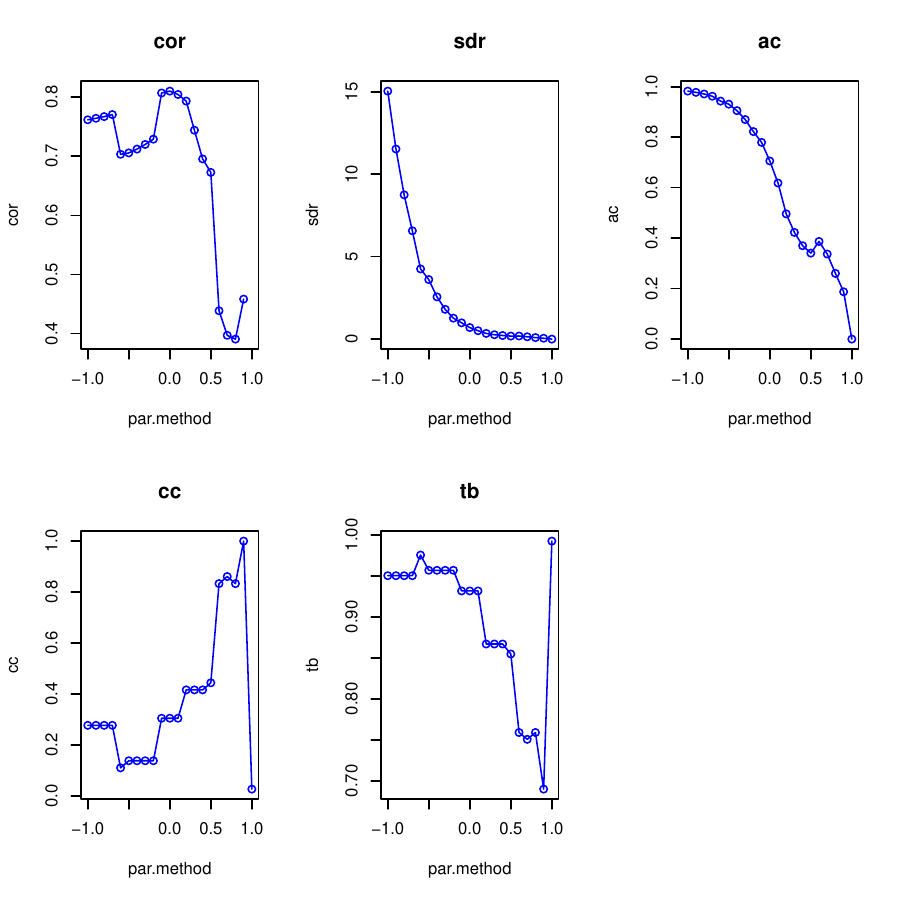}
\caption{\label{fig:descplotF} Descriptive measures obtained with the
\code{flexible} linkage method on the \code{UScitiesD} dataset. The best
cophenetic correlation coefficient (\code{cor}) is obtained when
\code{par.method} is close to $0$. Space distortion ratio (\code{sdr}),
agglomerative coefficient (\code{ac}) and tree balance (\code{tb}) decrease as
\code{par.method} increases, that is, moving from space-dilating to
space-contracting clustering structures. On the contrary, chaining coefficient
(\code{cc}) increases as \code{par.method} increases. Notice the exceptional
behavior observed when \code{par.method} is $1$, where \code{flexible} linkage
yields completely flat dendrograms.}
\end{figure}

It is interesting to know how the descriptive measures depend on the parameter
of the parametric linkage methods. Package \pkg{mdendro} provides the functions
\fct{descval} and \fct{descplot} for this task. For example, using the
\code{flexible} linkage method on the \code{UScitiesD} dataset (see
Figure~\ref{fig:descplotF}):
\begin{Schunk}
\begin{Sinput}
R> par(mfrow = c(2, 3))
R> measures <- c("cor", "sdr", "ac", "cc", "tb")
R> vals <- seq(from = -1, to = +1, by = 0.1)
R> for (m in measures)
+    descplot(UScitiesD, method = "flexible",
+             measure = m, par.method = vals,
+             type = "o",  main = m, col = "blue")
\end{Sinput}
\end{Schunk}

\subsubsection{Versatile linkage}

Package \pkg{mdendro} also implements another parametric linkage method named
versatile linkage \citep{Fernandez+Gomez:2020}. Substituting the arithmetic
means by generalized means, also known as power means, we can extend arithmetic
linkage to any finite power $p \neq 0$:
\begin{equation}
  \label{eq:versatile}
  D_{p}(X_{I},X_{J}) = \left( \frac{1}{|X_{I}||X_{J}|} \sum_{i \in I}
  \sum_{j \in J} |X_{i}||X_{j}| [D_{p}(X_{i},X_{j})]^{p} \right)^{1/p}\,,
\end{equation}
where $|X_{i}|$ and $|X_{j}|$ are the number of individuals in subclusters
$X_{i}$ and $X_{j}$, and $|X_{I}|$ and $|X_{J}|$ are the number of individuals
in clusters $X_{I}$ and $X_{J}$, i.e., $|X_{I}| = \sum_{i \in I} |X_{i}|$ and
$|X_{J}| = \sum_{j \in J} |X_{j}|$. Equation~\ref{eq:versatile} shows that
versatile linkage can be calculated using a combinatorial formula from the
distances $D_{p}(X_{i},X_{j})$ obtained during the previous iteration, in the
same way as the recurrence formula given in Equation~\ref{eq:Lance_Williams}.

Versatile linkage provides a way of obtaining an infinite number of AHC
strategies from a single formula, just changing the value of the power~$p$.
The decision of what power~$p$ to use can be taken in agreement with the type of
distance employed to measure the initial distances between individuals. For
instance, if the initial distances were calculated using a generalized distance
of order~$p$, then the natural AHC strategy would be versatile linkage with the
same power~$p$. However, this procedure does not guarantee that the dendrogram
obtained is the best one according to other criteria, e.g., cophenetic
correlation coefficient, space distortion ratio or tree balance (see
Section~\ref{subsec:descriptors}). Another possible approach consists in
scanning the whole range of parameters~$p$, calculate the preferred descriptors
of the corresponding dendrograms, and decide if it is better to substitute the
natural parameter~$p$ by another one. This is especially important when only the
distances between individuals are available, without coordinates for the
individuals, as is common in multidimensional scaling problems, or when the
distances have not been calculated using generalized means.

\begin{figure}[t!]
\centering
\includegraphics{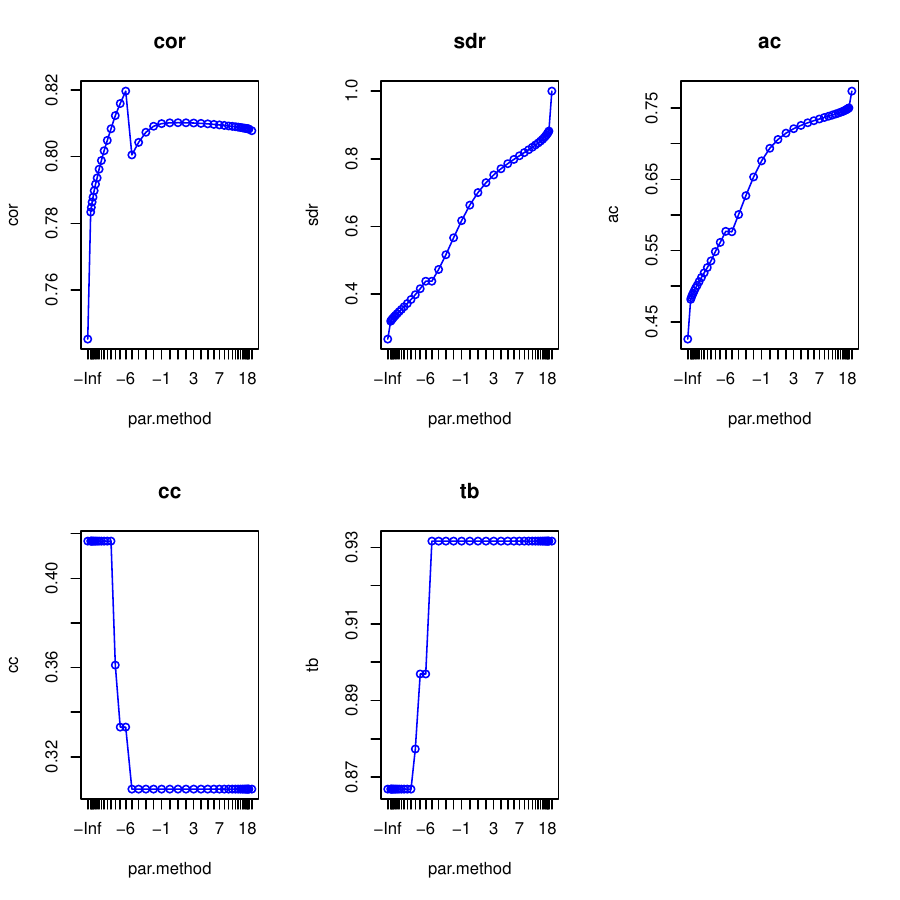}
\caption{\label{fig:descplotV} Descriptive measures obtained with the
\code{versatile} linkage method on the \code{UScitiesD} dataset. The best
cophenetic correlation coefficient (\code{cor}) is obtained when
\code{par.method} is close to $-6$. Space distortion ratio (\code{sdr}),
agglomerative coefficient (\code{ac}) and tree balance (\code{tb}) increase as
\code{par.method} increases, that is, moving from space-contracting to
space-dilating clustering structures. On the contrary, chaining coefficient
(\code{cc}) decreases as \code{par.method} increases. Notice that, unlike
\code{flexible} linkage, \code{sdr} never exceeds $1$ for \code{versatile}
linkage.}
\end{figure}

As in the case of \code{flexible} linkage, the parameter~$p$ of \code{versatile}
linkage is introduced using the \code{par.method} argument of the function
\fct{linkage}. Here, it is also interesting to know how the descriptors depend
on the parameter of this method (see Figure~\ref{fig:descplotV}):
\begin{Schunk}
\begin{Sinput}
R> par(mfrow = c(2, 3))
R> measures <- c("cor", "sdr", "ac", "cc", "tb")
R> vals <- c(-Inf, (-20:+20), +Inf)
R> for (m in measures)
+    descplot(UScitiesD, method = "versatile",
+             measure = m, par.method = vals,
+             type = "o",  main = m, col = "blue")
\end{Sinput}
\end{Schunk}

\textbf{Particular cases.} The generalized mean contains several well-known
particular cases, depending on the value of the power $p$. Some of them reduce
\code{versatile} linkage to the most commonly used methods, while others emerge
naturally as deserving special attention:
\begin{itemize}
  \item In the limit when $p \rightarrow -\infty$, \code{versatile} linkage
  becomes \code{single} linkage:
\begin{equation}
  D_{\min}(X_{I},X_{J}) = \min_{i \in I} \, \min_{j \in J} \,
  D_{\min}(X_{i},X_{j})\,.
\end{equation}
  \item In the limit when $p \rightarrow +\infty$, \code{versatile} linkage
  becomes \code{complete} linkage:
\begin{equation}
  D_{\max}(X_{I},X_{J}) = \max_{i \in I} \, \max_{j \in J} \,
  D_{\max}(X_{i},X_{j})\,.
\end{equation}
\end{itemize}
There are also three other particular cases that can be grouped together as
Pythagorean linkages. Therefore, in order to emphasize the existence of
different types of averages, we have preferred to rename average linkage as
\code{arithmetic} linkage:
\begin{itemize}
  \item When $p = +1$, the generalized mean is equal to the arithmetic mean and
    \code{arithmetic} linkage is recovered.
  \item When $p = -1$, the generalized mean is equal to the harmonic mean and
    \code{harmonic} linkage is obtained.
\begin{equation}
  D_{\mathrm{har}}(X_{I},X_{J}) = \left( \frac{1}{|X_{I}||X_{J}|} \sum_{i \in I}
  \sum_{j \in J} |X_{i}||X_{j}| [D_{\mathrm{har}}(X_{i},X_{j})]^{-1}
  \right)^{-1}\,.
\end{equation}
  \item In the limit when $p \rightarrow 0$, the generalized mean tends to the
    geometric mean and \code{geometric} linkage is obtained:
\begin{equation}
  D_{\mathrm{geo}}(X_{I},X_{J}) = \left( \prod_{i \in I} \prod_{j \in J}
  [D_{\mathrm{geo}}(X_{i},X_{j})]^{|X_{i}||X_{j}|} \right)^{1/(|X_{I}||X_{J}|)}
  \,.
\end{equation}
\end{itemize}
The correspondence between \code{versatile} linkage and the above mentioned
linkage methods is summarized in Table~\ref{tab:correspondence}.
\begin{table}[t!]
  \centering
  \begin{tabular}{lr}
    \hline
    & \code{versatile (par.method)} \\
	\hline
    \code{complete}   & \code{+Inf} \\
    \code{arithmetic} & \code{+1} \\
    \code{geometric}  & \code{0} \\
    \code{harmonic}   & \code{-1} \\
    \code{single}     & \code{-Inf} \\
	\hline
  \end{tabular}
  \caption{\label{tab:correspondence} Correspondence between \code{versatile}
  linkage and other linkage methods.}
\end{table}

\begin{figure}[t!]
\centering
\includegraphics{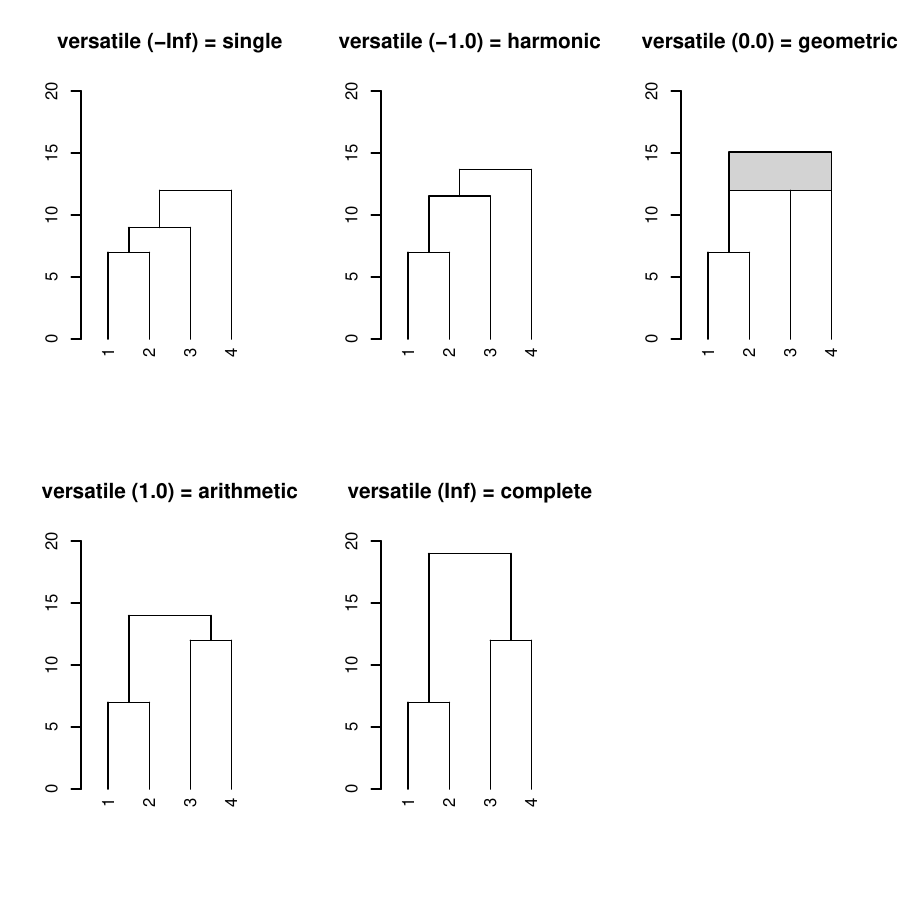}
\caption{\label{fig:versatile} Example of different dendrograms obtained as we
increase the versatile linkage parameter.}
\end{figure}

Let us show a small example in which we plot different dendrograms as we
increase the versatile linkage parameter, indicating the corresponding named
methods (see Figure~\ref{fig:versatile}):
\begin{Schunk}
\begin{Sinput}
R> d <- as.dist(matrix(c( 0,  7, 16, 12,
+                         7,  0,  9, 19,
+                        16,  9,  0, 12,
+                        12, 19, 12,  0), nrow = 4))
R> par(mfrow = c(2, 3))
R> vals <- c(-Inf, -1, 0, 1, Inf)
R> names <- c("single", "harmonic", "geometric", "arithmetic", "complete")
R> titles <- sprintf("versatile (%.1f) = %s", vals, names)
R> for (i in 1:length(vals)) {
+    lnk <- linkage(d, method = "versatile", par.method = vals[i], digits = 2)
+    plot(lnk, ylim = c(0, 20), cex = 0.6, main = titles[i])
+  }
\end{Sinput}
\end{Schunk}


\section{Comparison with other packages} \label{sec:comparison}

Except for the cases containing tied distances, the equivalences in
Table~\ref{tab:equivalences} hold between function \fct{linkage} in package
\pkg{mdendro}, function \fct{hclust} in package \pkg{stats} and function
\fct{agnes} in package \pkg{cluster}. Special attention must be paid to the
equivalence with methods \code{centroid} and \code{median} of function
\fct{hclust}, since these methods require the input distances to be squared
before calling \fct{hclust} and, consequently, the square root of its results
should be taken afterwards.
\begin{table}[t!]
  \centering
  \begin{tabular}{lll}
    \hline
    \fct{linkage} & \fct{hclust} & \fct{agnes}\\
	\hline
    \code{single} & \code{single} & \code{single}\\
    \code{complete} & \code{complete} & \code{complete}\\
    \code{arithmetic}, U & \code{average} & \code{average}\\
    \code{arithmetic}, W & \code{mcquitty} & \code{weighted}\\
    \code{geometric}, U/W & --- & ---\\
    \code{harmonic}, U/W & --- & ---\\
    \code{versatile}, U/W, $p$ & --- & ---\\
    --- & \code{ward} & --- \\
    \code{ward} & \code{ward.D2} & \code{ward}\\
    \code{centroid}, U & \code{centroid} & ---\\
    \code{centroid}, W & \code{median} & ---\\
    \code{flexible}, U, $\beta$ & --- & \code{gaverage}, $\beta$\\
    --- & --- & \code{gaverage}, $\alpha_{1}$, $\alpha_{2}$, $\beta$, $\gamma$\\
    \code{flexible}, W, $\beta$ & --- & \code{flexible}, $(1-\beta)/2$\\
    --- & --- & \code{flexible}, $\alpha_{1}$, $\alpha_{2}$, $\beta$, $\gamma$\\
	\hline
  \end{tabular}
  \caption{\label{tab:equivalences} Equivalences between functions
    \fct{linkage}, \fct{hclust} and \fct{agnes}. When relevant, weighted (W) or
    unweighted (U) versions of the linkage methods and the value for
    \code{par.method} are indicated.}
\end{table}

\begin{figure}[t!]
\centering
\includegraphics{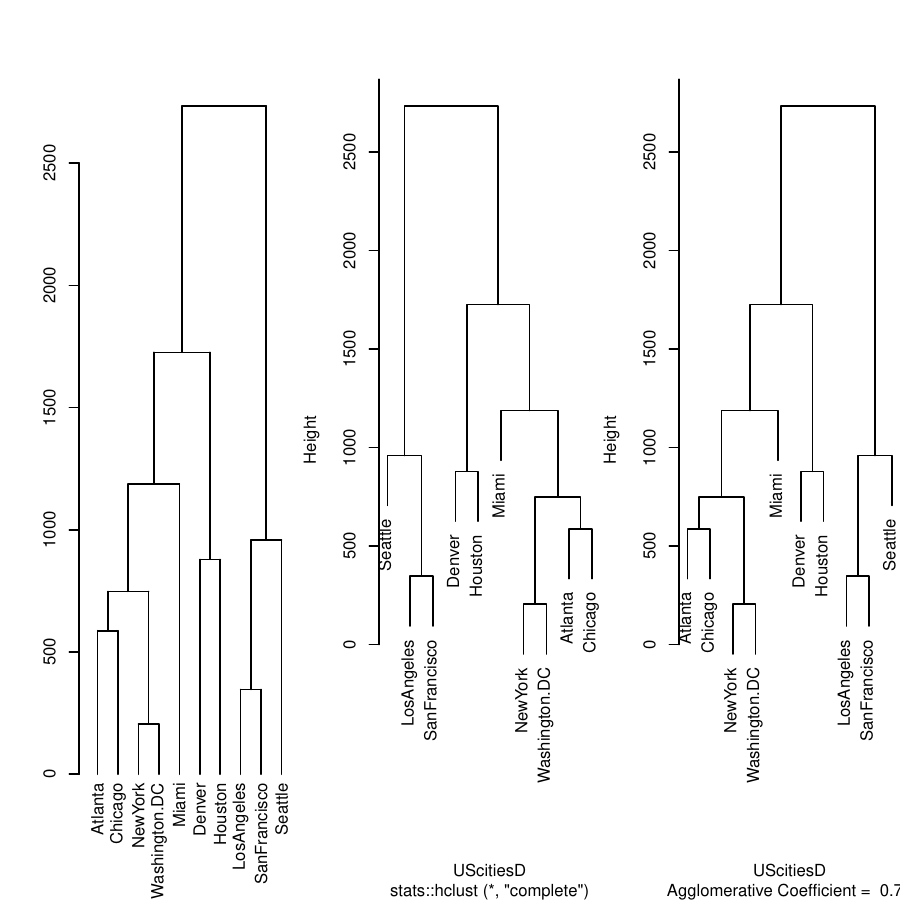}
\caption{\label{fig:comparison} Comparison of \code{complete} linkage on the
  \code{UScitiesD} dataset, using the functions \fct{linkage}, \fct{hclust} and
  \fct{agnes}.}
\end{figure}

For comparison, we can construct the same AHC using the functions \fct{linkage},
\fct{hclust} and \fct{agnes}, where the default plots just show some differences
in aesthetics (see Figure~\ref{fig:comparison}):
\begin{Schunk}
\begin{Sinput}
R> lnk <- mdendro::linkage(UScitiesD, method = "complete")
R> hcl <- stats::hclust(UScitiesD, method = "complete")
R> agn <- cluster::agnes(UScitiesD, method = "complete")
R> par(mar = c(5, 4, 4, 0), mfrow = c(1, 3))
R> plot(lnk)
R> plot(hcl, main = "")
R> plot(agn, which.plots = 2, main = "")
\end{Sinput}
\end{Schunk}

The cophenetic or ultrametric matrix is readily available as component
\code{coph} of the returned \class{linkage} object, and coincides with those
obtained using the functions \fct{hclust} and \fct{agnes}:
\begin{Schunk}
\begin{Sinput}
R> hcl.coph <- cophenetic(hcl)
R> all(lnk$coph == hcl.coph)
\end{Sinput}
\begin{Soutput}
[1] TRUE
\end{Soutput}
\begin{Sinput}
R> agn.coph <- cophenetic(agn)
R> all(lnk$coph == agn.coph)
\end{Sinput}
\begin{Soutput}
[1] TRUE
\end{Soutput}
\end{Schunk}

The coincidence also applies to the cophenetic correlation coefficient and the
agglomerative coefficient, with the advantage that function \fct{linkage} has
both of them already calculated:
\begin{Schunk}
\begin{Sinput}
R> hcl.cor <- cor(UScitiesD, hcl.coph)
R> all.equal(lnk$cor, hcl.cor)
\end{Sinput}
\begin{Soutput}
[1] TRUE
\end{Soutput}
\begin{Sinput}
R> all.equal(lnk$ac, agn$ac)
\end{Sinput}
\begin{Soutput}
[1] TRUE
\end{Soutput}
\end{Schunk}

\begin{figure}[t!]
\centering
\includegraphics{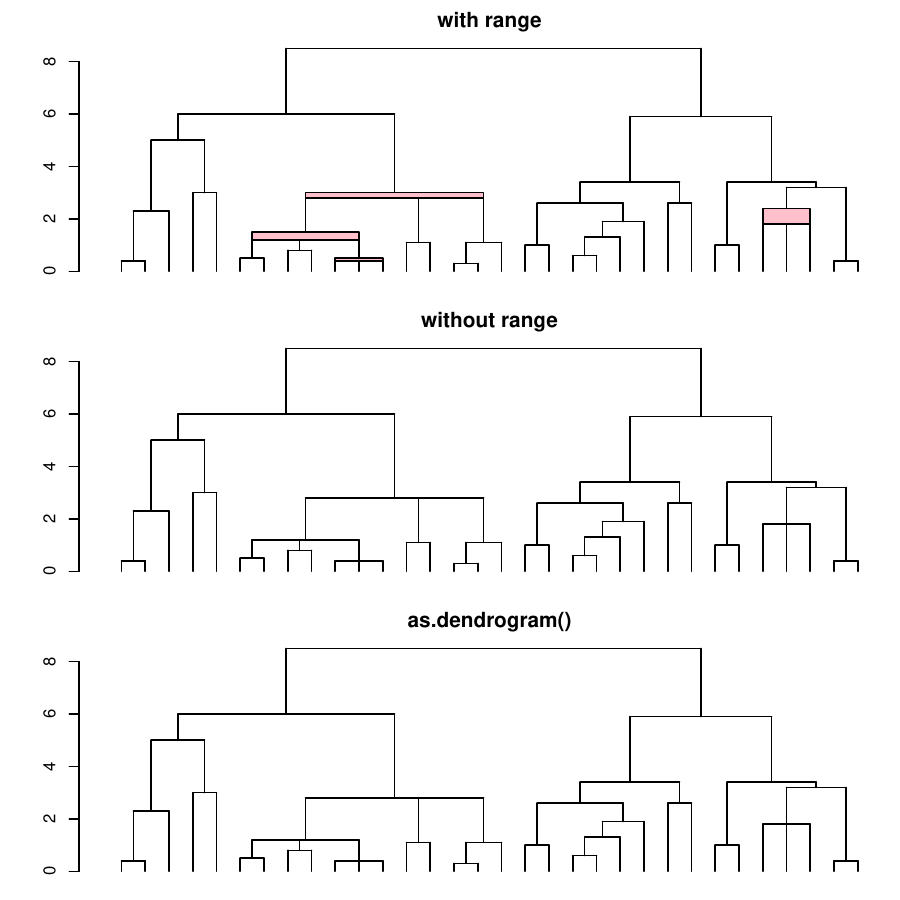}
\caption{\label{fig:asdendrogram} Example of plotting a dendrogram with ties
drawing a range rectangle for tied distances, hiding it, and plotting the result
returned by function \fct{as.dendrogram}.}
\end{figure}

To enhance usability and interoperability, class \class{linkage} includes method
\fct{as.dendrogram} for class conversion. In the example shown in
Figure~\ref{fig:comparison}, converting to class \class{dendrogram} the
objects returned by the functions \fct{linkage}, \fct{hclust} and \fct{agnes},
we can see that all three dendrograms are structurally equivalent. Since class
\class{dendrogram} can only handle binary edges, function \fct{as.dendrogram}
works by converting tied distances into consecutive binary edges at the same
height, thus having the same visual effect as having a single edge with more
than two children. Using the same example as in Figure~\ref{fig:ranges}, which
contains tied distances, we can observe that plotting a dendrogram with ties as
returned by function \fct{as.dendrogram} is the same as plotting it hiding the
range rectangles for tied distances (see Figure~\ref{fig:asdendrogram}):
\begin{Schunk}
\begin{Sinput}
R> cars <- round(dist(scale(mtcars)), digits = 1)
R> lnk <- linkage(cars, method = "complete", group = "variable")
R> par(mfrow = c(3, 1), mar = c(1, 4, 2, 0))
R> plot(lnk, col.rng = "pink", main = "with range", leaflab = "none")
R> plot(lnk, col.rng = NULL, main = "without range", leaflab = "none")
R> plot(as.dendrogram(lnk), main = "as.dendrogram()", leaflab = "none")
\end{Sinput}
\end{Schunk}

\begin{figure}[t!]
\centering
\includegraphics{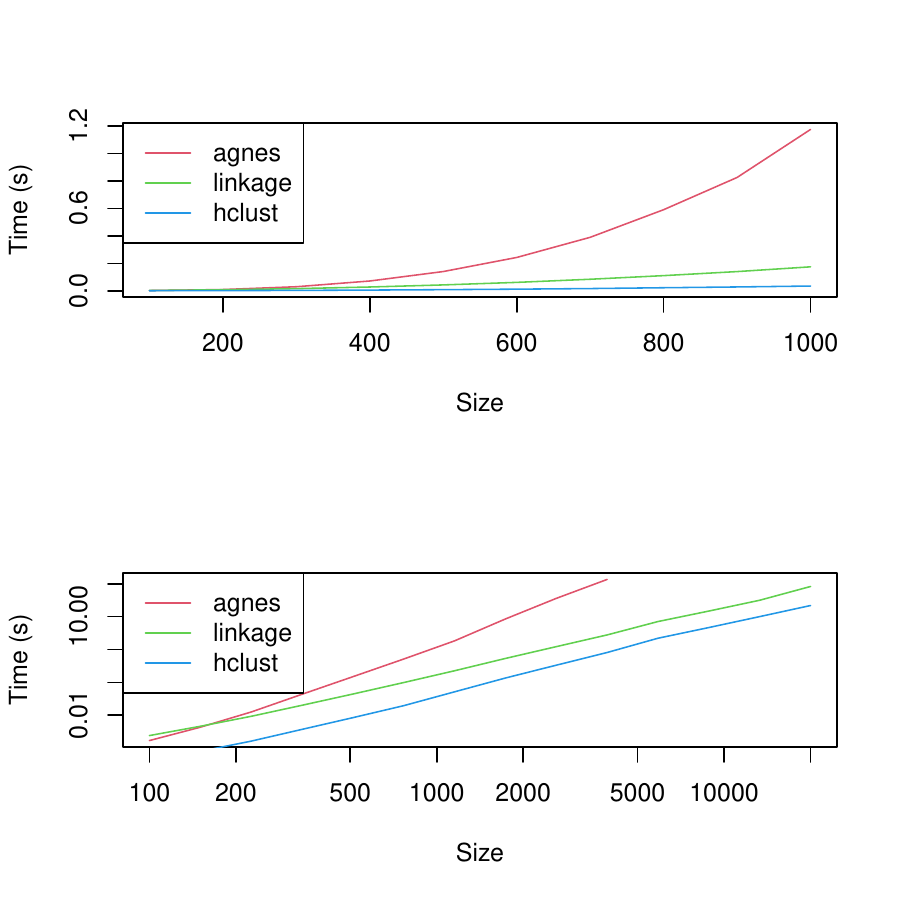}
\caption{\label{fig:efficiency} Comparison of the computational efficiency of
functions \fct{agnes}, \fct{linkage} and \fct{hclust}, both in linear scale
(top) and in log-log scale (bottom). The results are averages of the time taken
to compute the dendrogram by each function, calculated over 20~random distance
matrices for each size and method (single, complete, arithmetic and ward), and
also averaging over the methods. The slope of the lines in the log-log plot
indicates the different exponents of the cost.}
\end{figure}

The computational efficiency of functions \fct{linkage}, \fct{hclust} and
\fct{agnes} is compared in Figure~\ref{fig:efficiency}, where it can be observed
that the time cost of functions \fct{linkage} and \fct{hclust} is quadratic
(exponents 1.99 and 2.12 respectively), whereas that of function \fct{agnes} is
cubic (exponent 3.12).

\begin{figure}[t!]
\centering
\includegraphics{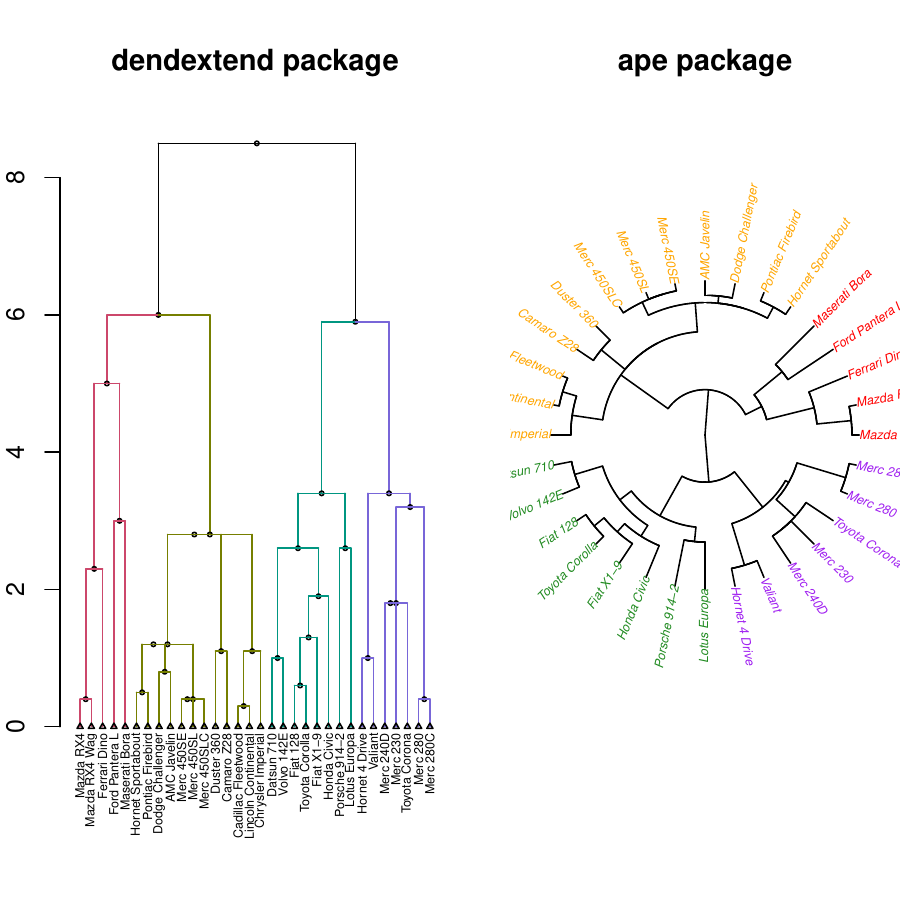}
\caption{\label{fig:packages} Converting objects of class \class{linkage} using
  the function \fct{as.dendrogram}, one can take advantage of other dendrogram
  plotting packages, such as \pkg{dendextend} and \pkg{ape}.}
\end{figure}

Plots including ranges are only available if the user directly employs the
\fct{plot.linkage} function from package \pkg{mdendro}. Anyway, the user may
still take advantage of other dendrogram plotting packages, such as
\pkg{dendextend} \citep{Galili:2015} and \pkg{ape} \citep{Paradis+Schliep:2019}
(see Figure~\ref{fig:packages}):
\begin{Schunk}
\begin{Sinput}
R> par(mar = c(5, 2, 4, 0), mfrow = c(1, 2))
R> cars <- round(dist(scale(mtcars)), digits = 1)
R> lnk <- linkage(cars, method = "complete")
R> lnk.dend <- as.dendrogram(lnk)
R> plot(dendextend::set(lnk.dend, "branches_k_color", k = 4),
+       main = "dendextend package",
+       nodePar = list(cex = 0.4, lab.cex = 0.5))
R> lnk.hcl <- as.hclust(lnk)
R> pal4 <- c("red", "forestgreen", "purple", "orange")
R> clu4 <- cutree(lnk.hcl, k = 4)
R> plot(ape::as.phylo(lnk.hcl),
+       type = "fan",
+       main = "ape package",
+       tip.color = pal4[clu4],
+       cex = 0.5)
\end{Sinput}
\end{Schunk}

\begin{figure}[t!]
\centering
\includegraphics{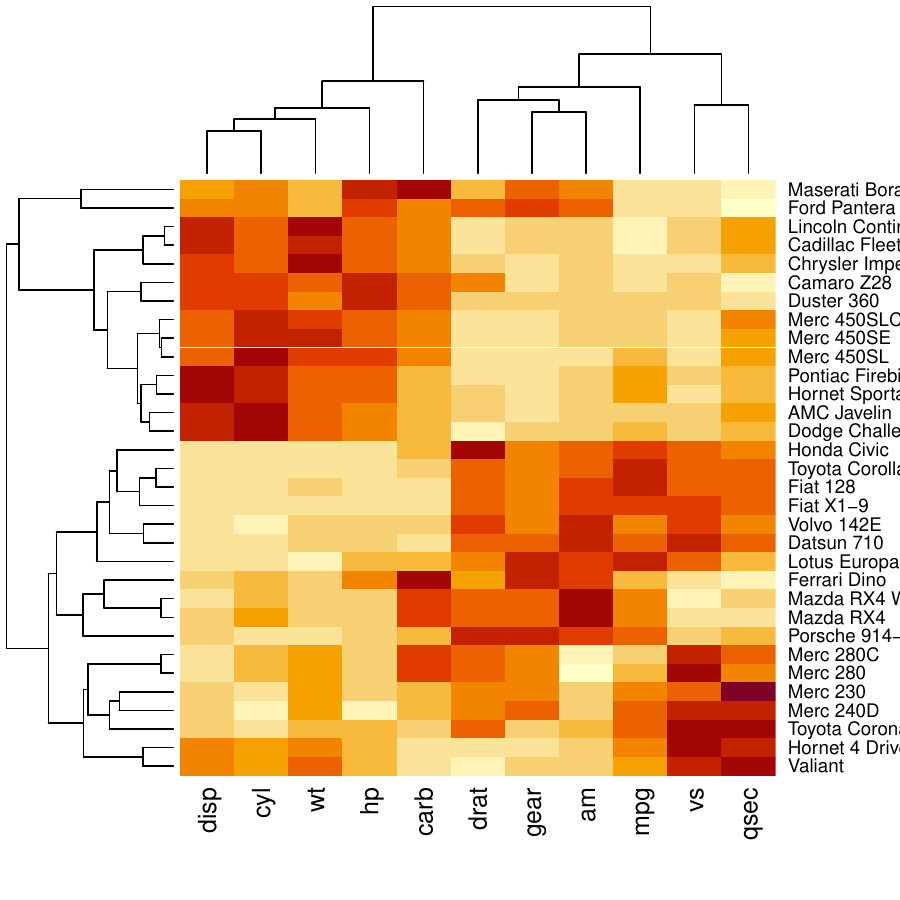}
\caption{\label{fig:heatmap} Example of heatmap constructed using the function
  \fct{linkage}.}
\end{figure}

And users can also use function \fct{linkage} to plot heatmaps containing
multidendrograms (see Figure~\ref{fig:heatmap}):
\begin{Schunk}
\begin{Sinput}
R> heatmap(scale(mtcars), hclustfun = linkage)
\end{Sinput}
\end{Schunk}

\begin{figure}[t!]
\centering
\includegraphics{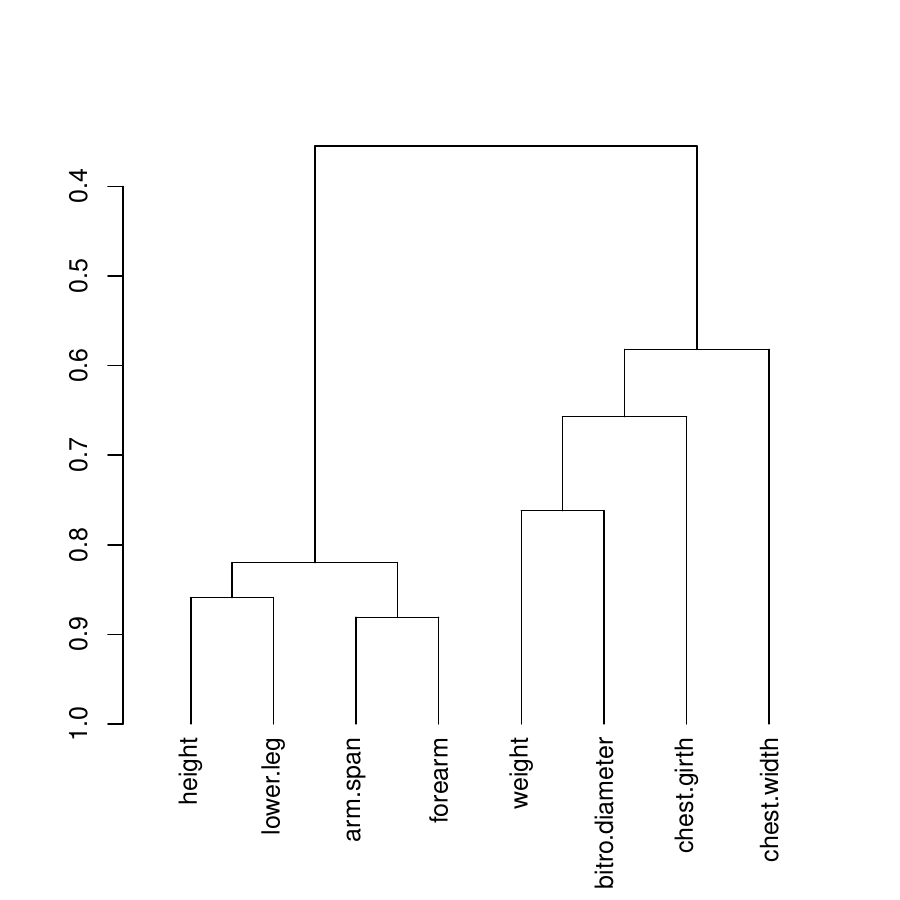}
\caption{\label{fig:similarity} Example of a dendrogram constructed from a
  matrix of nonnegative correlations, i.e., directly using similarities instead
  of distances.}
\end{figure}

In addition, it is possible to work directly with similarity data without having
to convert them to distances, provided they are in the range [0.0, 1.0]. The
value 1.0 is selected here as the origin value for similarity dendrograms since
it is a value commonly used to measure the similarity between an object and
itself; otherwise, the original similarities could be normalized to fulfill this
requirement. A typical example would be a matrix of nonnegative correlations
(see Figure~\ref{fig:similarity}):
\begin{Schunk}
\begin{Sinput}
R> sim <- as.dist(Harman23.cor$cov)
R> lnk <- linkage(sim, type.prox = "sim")
R> plot(lnk)
\end{Sinput}
\end{Schunk}

It is important to remark that in the literature there exist several types of
extensions of AHC for similarity data that are consistent with the standard
distance formulation, such as extensions based on kernel methods
\citep{Ah-Pine:2018}. The extension of function \fct{linkage} to similarities
is based on two straightforward changes: a) in the agglomeration process, join
the clusters with highest similarity (instead of those at lowest distance);
b) define how to calculate the similarity between clusters. For single
linkage, the similarity between clusters is defined as the maximum similarity
between their components; for complete linkage, the similarity between clusters
is defined as the minimum similarity between their components; and for the rest
of the methods, the similarity between clusters is calculated exactly in the
same way that the distance between clusters, i.e.,
Equations~\ref{eq:Lance_Williams} and~\ref{eq:versatile} are unchanged.


\section{Summary and discussion} \label{sec:summary}

\pkg{mdendro} is a simple yet powerful \proglang{R} package to make hierarchical
clusterings of data. It implements a variable-group algorithm for AHC that
solves the nonuniqueness problem found in pair-group algorithms. This problem
consists in obtaining different hierarchical clusterings from the same matrix of
pairwise distances, when two or more shortest distances between different
clusters are equal during the agglomeration process. In such cases, selecting a
unique clustering can be misleading. Software packages that do not ignore this
problem fail to adopt a common standard with respect to ties, and many of them
simply break ties in any arbitrary way.

Package \pkg{mdendro} computes dendrograms grouping more than two clusters at
the same time when ties occur. It includes and extends the functionality of
other reference packages in several ways:
\begin{itemize}
  \item Native handling of both distance and similarity matrices.
  \item Calculation of variable-group multifurcated dendrograms, which solve the
    nonuniqueness problem of AHC when there are tied distances.
  \item Implementation of the most common AHC linkage methods: single linkage,
    complete linkage, average linkage, centroid linkage and Ward's method.
  \item Implementation of two parametric linkage methods: $\beta$-flexible
    linkage and versatile linkage. The latter leads naturally to the definition
    of two new linkage methods: harmonic linkage and geometric linkage.
  \item Implementation of both weighted and unweighted forms for the previous
    linkage methods.
  \item Calculation of the cophenetic (or ultrametric) matrix.
  \item Calculation of five descriptive measures for the resulting dendrogram:
    cophenetic correlation coefficient, space distortion ratio, agglomerative
    coefficient, chaining coefficient and tree balance.
  \item Plots of the descriptive measures for the parametric linkage methods.
\end{itemize}

Although ties need not be present in the initial proximity data, they may arise
during the agglomerative process. For this reason, and given that the results of
the variable-group algorithm coincide with those of the pair-group algorithm
when there are no ties, we recommend to directly use package \pkg{mdendro}. With
a single action one knows whether ties exist or not, and additionally the
subsequent hierarchical clustering is obtained.


\section*{Computational details}

The results in this paper were obtained using
\proglang{R}~4.4.1 with the
\pkg{mdendro}~2.2.1 package. \proglang{R} itself
and all packages used are available from the Comprehensive \proglang{R} Archive
Network (CRAN) at \url{https://CRAN.R-project.org/}.

\section*{Acknowledgments}

This work was supported by Ministerio de Ciencia e Innovaci{\'o}n
(PID2021-124139NB-C22, PID2021-128005NB-C21, RED2022-134890-T and
TED2021-129851B-I00), Generalitat de Catalunya (2021SGR-633) and
Universitat Rovira i Virgili (2021PFR-URV-100 and 2022PFR-URV-56).




\end{document}